\documentclass[prd,aps,floats,twocolumn,nofootinbib]{revtex4-1}
\usepackage{slashed}
\usepackage{mathtools}
\usepackage{amsfonts}
\usepackage{amssymb}
\usepackage{epsfig}

\usepackage{xcolor}

\begin{document}

\newcommand{\Sim}[1]{\textcolor{red}{\bf #1}}

\newcommand{\m}[1]{\mathcal{#1}}
\newcommand{\nn}{\nonumber}
\newcommand{\ph}{\phantom}
\newcommand{\eps}{\epsilon}
\newcommand{\be}{\begin{equation}}
\newcommand{\ee}{\end{equation}}
\newcommand{\bea}{\begin{eqnarray}}
\newcommand{\eea}{\end{eqnarray}}
\newtheorem{conj}{Conjecture}


\title{Quantum cosmology of a dynamical Lambda}
\date{\today}

\author{Jo\~{a}o Magueijo}
\email{j.magueijo@imperial.ac.uk}
\affiliation{Theoretical Physics Group, The Blackett Laboratory, Imperial College, Prince Consort Rd., London, SW7 2BZ, United Kingdom}
\author{Tom Zlosnik}
\email{zlosnik@fzu.cz}
\affiliation{CEICO, Institute of Physics of the Czech Academy of Sciences, Na Slovance 1999/2, 182 21, Prague}
\author{Simone Speziale}
\email{simone.speziale@gmail.com}
\affiliation{Centre de Physique Theorique,
Aix Marseille Univ., Univ. de Toulon, CNRS, Marseille, France}

\begin{abstract}
By allowing torsion into the gravitational dynamics one can promote the cosmological constant, $\Lambda$, to a dynamical variable in a class of quasi-topological theories. In this paper we perform a mini-superspace quantization of these theories in the connection representation. If $\Lambda$ is kept fixed, the solution is a delta-normalizable
version of the Chern-Simons (CS) state, 
which is the dual of the Hartle and Hawking and Vilenkin wave-functions.  
We find that the CS state solves the Wheeler-De Witt equation also if 
$\Lambda$ is rendered dynamical by an Euler quasi-topological invariant, {\it in the parity-even branch of the theory}. In the absence of an infra-red (IR) cut-off, the CS state suggests 
the marginal probability $P(\Lambda)=\delta(\Lambda)$. Should there be an IR cutoff (for whatever reason) the probability is sharply peaked at the cut off. In the parity-odd branch, however, 
we can still find the CS state as a particular (but not most general) solution, but further work is needed to sharpen the predictions.
For the theory based on the Pontryagin invariant (which only has a parity-odd branch) the CS wave function no longer is a solution to the constraints. We find the most general solution in this case, which again leaves room for a range of predictions
for $\Lambda$. 
\end{abstract}

\maketitle

\section{Introduction}
The cosmological constant problem is a nuisance that has been with us since its introduction by Einstein in the aftermath of General Relativity's (GR) proposal. And yet we have also learned a fair amount about fundamental physics
from this alleged problem. Indeed the Lambda ``problem'' has metamorphosed into so many different forms, that 
it can be said that there is not one, but a multitude of distinct, often independent problems (for two excellent reviews, some thirty years apart, see~\cite{weinberg,padilla}).  Thus, it will be inevitable that in this paper we will fail
to address the problem in the incarnation considered to be  the ``real'' one by some authors. We do not take a partisan view on the matter, but hope that  our paper may be a useful contribution to the overall discussion
of the meaning of the cosmological constant. 

It was noted long ago that one gains new perspectives on the cosmological constant problem if the epithet ``constant'' becomes a misnomer. Notably, Hawking suggested~\cite{hawking} that the introduction of a gauge 3-form would promote $\Lambda$ to a dynamical variable, only forced to be a constant on-shell. This may lead to a solution to the problem within the path integral formalism (see~\cite{hawking}  but also~\cite{weinberg}, Sec. VIII). We acknowledge this work as our main source of inspiration, but in contrast to Hawking's work, in this paper $\Lambda$ will not be forced to constancy by the classical equations of motion, with significant consequences. Our work  is based on a class of theories of gravity~\cite{alex1,alex2,MZ} minimally extending the Einstein-Cartan (EC) formulation of General Relativity so as to accommodate a classically variable $\Lambda$ through a balancing torsion. 
The purpose of this paper is to derive the quantum cosmology of these theories. 
In the process we may accidentally revitalize the Chern-Simons (CS) state~\cite{kodama}  as a proposal for the wave function of the Universe (see~\cite{randono1,randono2} for other attempts to improve upon earlier work~\cite{lee1,lee2,witten}; also~\cite{CSHHV}).

The theories envisaged in~\cite{alex1,MZ} are {\it quasi}-topological theories based on the Euler and Pontryagin invariants multiplied by $1/\Lambda$. The topological nature of the invariants is disrupted by $\Lambda$ being elevated to a dynamical variable,
but restored for solutions with constant $\Lambda$. Interesting cosmological consequences were found in~\cite{alex2,MZ}, 
with the remarkable result that the dynamics has two entirely separate branches, with different constraints, continuous symmetries and number of degrees of freedom, depending on whether or not parity violation is permitted~\cite{MZ}. Parity-odd solutions can be consistent with homogeneity and isotropy because 
of torsion: the so-called Cartan spiral staircase~\cite{spiral}. Whether this structure is switched on or off leads to different Hamiltonian dynamics and sets of constraints.

While the cosmological phenomenology of these theories has the potential to deviate considerably from the standard model for some parts of the model's parameter space ~\cite{alex2,MZ,alex-grav} (though see also \cite{Barrientos:2019awg}), the point remains that they could form interesting quantum cosmology models. Their mini-superspace (MSS) can be used to investigate questions of Quantum Gravity.   
We find that 
for the dynamical $\Lambda$ theory based on the Euler topological invariant the general solution is an
adaptation and generalization of the CS wave, which
is always delta-function normalizable. At the kinematical level, 
it implies
$P(\Lambda)=\delta(\Lambda)$ for the marginal probability of $\Lambda$. In the parity violating branch of the theory
the situation is more complicated and dependent on the boundary conditions imposed for the parity-odd component of the torsion. 
The wave function  can still be related to 
a generalization of the CS state, but now this can be multiplied by an amplitude with a specific functional
dependence. For the theory based on the Pontryagin invariant the generalization of the CS state no longer provides a solution,
but we are able to derive the most general solution, which once again leave greater room for prediction about Lambda. 

The plan of this paper is as follows.  In Section~\ref{classth} we review the classical theory and its MSS reduction (conjugate variables and constraints) in the case where the Pontryagin term is ignored. In Section~\ref{ECth} we consider pure EC theory (obtained from ours by forcing $\Lambda$ to be a fixed parameter $\Lambda_0$), deriving its most obvious solution: the CS wave-function. This is shown to be the Fourier dual of other wave-functions considered before, namely the Hartle-Hawking and Vilenkin wave-functions, as studied in more detail elsewhere~\cite{CSHHV}. 

Then, in Section~\ref{EulerPEth} we examine the theory based on the quasi-Euler term focusing on its parity-even branch. This is the simplest theory in which $\Lambda$ can vary classically, and we find that the solution is a straightforward generalization of the CS state (one where $\Lambda_0$ is replaced by coordinate $\Lambda$). In Section~\ref{probL} we investigate the implications for Lambda, face value, to conclude that the state implies $P(\Lambda)=\delta(\Lambda)$. Should there be IR and UV cutoffs, the probability is heavily peaked on the IR cut off. The matter, however, depends on the choice of inner product. 

The situation is less clear in the parity-violating branch of these theories. In Section~\ref{EulerPOth} we find that the generalized CS state (with $\Lambda_0$ replaced by a coordinate, and the CS functional including the parity-odd contribution from the connection)
still provides a particular solution to the quantum constraints; however, the most general solution allows for the CS wave function to be multiplied by an amplitude with very specific functional dependence. This allows for a greater range of predictions. 

The problem is again very different for the theory based on the Pontryagin invariant (which only has a parity-odd branch).
This is studied in Section~\ref{PontTH}, where in Section~\ref{PontHam} we extend the classical analysis reviewed in Section~\ref{classth} to cover this case. In Section~\ref{PontWdW} we find that the same generalized CS wave function no longer is a solution to the constraints. We find the most general solution in this case, finding room for a large range of predictions
for $\Lambda$. In Section~\ref{comparison} we compare our results to other results in the literature where $\Lambda$ is considered dynamical and in Section~\ref{matter} we briefly discuss coupling our model to matter and some of the challenges this presents. Finally we present our conclusions.


\section{The classical theory and its minisuperspace}\label{classth}
Our classical theory has gravitational action:
\be\label{Sg}
S^g[e,\omega,\Lambda]=\frac{\kappa}{2} (S_{Pal}+
S_{Eul}+ S_{Pont})
\ee
with
\bea
S_{EC}&=&\int  \epsilon_{ABCD}\left( e^A e^B R^{CD} -\frac{\Lambda}{6} e^A e^Be^Ce^D\right)  \label{palatini},\\
S_{Eul}&=&-\frac{3}{2}\int \frac{1}{\Lambda} \epsilon_{ABCD}R^{AB}R^{CD} ,\label{euler}\\
S_{Pont}&=&-\frac{3}{\gamma}\int \frac{1}{\Lambda} R^{AB}R_{AB},
\eea
where we describe gravity by the scalar $\Lambda$, the co-tetrad  $e^{A}=e^{A}_{\mu}dx^{\mu}$, and the spin-connection $\omega^{A}_{\ph{A}B\mu}dx^{\mu}$ and multiplication of differential forms is via the wedge product unless otherwise stated\footnote{The action (\ref{Sg}) is in the first-order formalism of gravity where gravitational connection and frame field or metric are regarded as independent objects. This remains a relatively unexplored setting for the exploration of modified theories of gravity beyond General Relativity. See \cite{Calcagni:2009xz,Alexander:2012ab,BeltranJimenez:2017tkd,Alexandrov:2019dxm} for interesting recent proposals.}. The quantity $R^{AB}=d\omega^{AB}+\omega^{A}_{\ph{A}C}\omega^{CB}$ is the curvature two form, $S_{EC}$ is the Einstein-Cartan action (with a Lambda term), and $S_{Eul}$ and $S_{Pont}$ are the Euler and Pontryagin quasi-topological terms (their topological nature disrupted by the potential non-constancy of $\Lambda$).  We choose $\kappa=1/(16\pi G_N)$ to make the normalisation consistent with the Einstein-Hilbert action in the appropriate limit\footnote{This changes the notation by correcting the normalization used in~\cite{alex1,alex2,MZ} by a factor of 1/2.
Note that the definition of stress-energy 3-forms in~\cite{alex1,alex2,MZ} ajusts to the different normalization used for $S_g$,
so it is all consistent. Likewise, for the discussion in Appendix II in~\cite{MZ}.}.
We have suppressed the Nieh-Yan topological invariant, since we will keep the action explicitly real throughout, and used $\gamma$ to parameterize the Pontryagin term (noting that it should not be confused with the Immirzi parameter; see proviso in~\cite{MZ} and also 
Appendix I of this paper).

We then reduce the action to MSS assuming first no Pontryagin term\footnote{It turns out that the Pontryagin case is very different, and indeed has not been studied before.} ($\gamma\rightarrow\infty$), 
referring the reader to~\cite{MZ} for full details. 
Here $a$ is the expansion factor, $k=0,\pm 1$ is the spatial curvature, $b$ is the parity-even component of the connection (classically 
related to the expansion rate and the parity-even part of the torsion, $T(t)$, according to
$b=\dot a+ aT$) and $c$ is the parity-odd part of the torsion. As explained in~\cite{MZ} (see Appendix I
and the start of Section V), the reduced action may be written as:
\begin{align}
\label{Faction2}
S&=3\kappa V_c \int dt \bigg(  2a^{2} \dot{b}+\Pi \frac{d\Lambda^{-1}}{dt} + 2Na {\cal H}+V{\cal V}\bigg)\\
{\cal H} &= b^{2}+k-c^{2}-\frac{\Lambda}{3}a^{2}\\
{\cal V} &= \frac{\Pi}{6}-\bigg(\frac{b^{3}}{3}+b\bigg(k-c^{2}\bigg)\bigg)
\end{align}
where $\Pi$ is the moment conjugate to $1/\Lambda$, subject to a new constraint ${\cal V}$, in addition to the Hamiltonian constraint ${\cal H}$.  This is the action found in Eq. 70 of Ref.~\cite{MZ}, with coupling $\kappa$ and spatial integration restored. The latter produces a factor of $6V_c$, with $V_c=\int d^3 x$
an integral in comoving variables. 
In quantum cosmology this is usually taken over the whole space, assumed
to be compact, typically with $k=1$. In that case $V_c=2\pi^2$ (whole 3-sphere). 
Elsewhere, we will explore other possibilities, such as,  
topologically non-trivial versions of $k=0,-1$ with finite $V_c$, of sub-regions of the whole space. 
Note that the overall factors in action (\ref{Faction2}) are irrelevant for pure gravity in the classical theory, but 
{\it not} for the quantum theory.  

The Poisson brackets can be read off from (\ref{Faction2}) and are:
\bea
\{b,a^2\}&=&\frac{1}{6\kappa V_c}\label{PBba}\\
\{\Lambda^{-1},\Pi\}&=&\frac{1}{3\kappa V_c}.\label{PBLPi}
\eea
The constraints are always first class~\cite{MZ}, so the Dirac and the Poisson bracket are identical.
Quantization, therefore, implies commutation relations:
\bea
\left[\hat b,\hat{ a^2}\right]&=&\frac{i l_P^2}{3 V_c}\label{com1}\\
\left[\hat \Lambda^{-1},\hat{ \Pi}\right]&=&\frac{2 i l_P^2}{3 V_c}\label{com2}
\eea
where $l_P$ is the reduced Planck length, $l_P=\sqrt{8\pi G_N \hbar}$ (note that in some literature this is defined as
$l_P=\sqrt{16 \pi G_N \hbar}$ explaining some disparity in factors). 

The quantum constraint equations can now be written down on a case by case basis. Note that overall factors of $a$ (such as the one
appearing in the Hamiltonian constraint) may be factored out but only with assumptions about ordering.
This will be important in making contact with standard results in quantum cosmology, typically using the $a$ representation, instead of the $b$ representation used here.\footnote{ The connection representation is also the starting point for loop quantum cosmology \cite{Ashtekar:2011ni}. The  difference there is the use of a different quantization of the algebra, motivated by loop quantum gravity, where wave functions depend on the connection only through their holonomies. }

\section{Einstein-Cartan theory and the Chern-Simons state}\label{ECth}
To illustrate our methods we first solve the problem 
in the simpler case of Einstein-Cartan theory, obtained by forcing $\Lambda$ to be the fixed parameter
$\Lambda_0$. 
It turns out that in this context our results are nothing but standard quantum cosmology in a 
Fourier-dual representation, as studied in more detail in~\cite{CSHHV}.

Within Einstein-Cartan theory, the reduced action, instead of (\ref{Faction2}), is:
\bea\label{ECaction}
S
&=&3\kappa V_c \int dt \bigg(  2a^{2} \dot{b}
+ 2Na\bigg(b^2+k-\frac{\Lambda_{0}}{3}a^{2}\bigg) \bigg).
\eea

In the connection representation, where $\psi=\psi(b)$, the single commutator 
(\ref{com1}) is satisfied by:
\be
\hat a^2=-\frac{i l_P^2}{3 V_c}\frac{d}{db}.
\ee
Hence, a possible expression the Hamiltonian constraint equation takes the form:
\be\label{wdweq}
\hat{\cal H}\psi=\bigg(\frac{i \Lambda_{0} l_P^2}{9 V_c}\frac{d}{db} + k+b^2 \bigg)\psi=0.
\ee
As already explained, we have factored out an $a$ in this rendition of the Hamiltonian constraint, so an assumption on ordering has been made.

This equation integrates trivially to:
\be\label{kod0}
\psi_K={\cal N} \exp{\bigg(i\frac{ 9 V_c}{\Lambda_{0} l_P^2} \left(\frac{b^3}{3}+kb\right)\bigg)}
\ee
(up to a constant phase).
As explained in  more generality and detail in Appendix~\ref{CSapp}, this is the CS wave-function applied to this theory (EC) and reduction to Friedmann-Robertson-Walker (FRW) spacetime symmetry. 
This is the first instance in this paper where we find that an extension or straightforward adaptation of the Chern-Simons state appears as a solution to our theories.  Invariably the state found will be an exponential with an imaginary phase, thereby bypassing many of the criticisms levelled upon this state, as we will discuss later. 

Not only is this state non-pathological, but also it is nothing but the
Fourier dual of well known wave functions of the Universe (usually expressed in the representation
diagonalizing $\hat a$ instead of $\hat b$), such as the Hartle and Hawking~\cite{HH,Vilenkin} and the Vilenkin~\cite{vil0,Vilenkin}
proposals. This is examined in detail in~\cite{CSHHV}; here we merely sketch the starting point of the argument, for the sake 
of the rest of this paper.

Recall that the Einstein-Hilbert actions for mini-superspace is:
\be\label{EHaction}
S=6\kappa V_c\int dt \left(\ddot a a^2 + \dot a^2a +ka -\frac{\Lambda_0}{3}a^3\right).
\ee
This results in the action used in standard treatments (for a good review see~\cite{Vilenkin}), after one 
integrates the first term by parts, and specialises to 
$k=1$, and integrates over the whole 3-sphere ($V_c=2\pi^2$).
With some adaptation of notation\footnote{To bridge notation notice that $6\kappa V_c=1/2\sigma^2$ as defined in~\cite{Vilenkin}. Also note that $\Lambda_0$ is defined with an extra factor of 1/3 there.} a Hamiltonian analysis of
this action leads to the Wheeler-De Witt (WdW) equation:
\be\label{WdWa}
\left[\frac{d^2}{da^2}+\frac{\alpha}{a}\frac{d}{da}-U(a)\right]\psi=0
\ee
where $\alpha$ is a parameter encoding an ordering ambiguity and 
\be
U(a)=4\left(\frac{3V_c}{l_P^2}\right)^2 a^2\left(k-\frac{\Lambda_0}{3}a^2\right)
\ee
is the effective potential. 

It is not difficult to see that actions (\ref{EHaction}) and (\ref{ECaction}) are equivalent if
there is no torsion (an assumption we will have to drop for the rest of this paper). Setting $ b=\dot a/N$ (since $T=0$)
in (\ref{EHaction}) promptly leads to (\ref{ECaction}), showing that the the conjugate pairs and Hamiltonian 
constraint are the same in both theories. Thus, their difference in aspect reflects merely the fact that in (\ref{wdweq})
we used the $b$ representation. Had we used the $a$ representation in the EC theory, then 
(\ref{com1}) would have implied:
\be
\hat b=\frac{i l_P^2}{3 V_c}\frac{d}{d(a^2)}
\ee
so that $\hat {\cal H}\psi=0$ (instead of (\ref{wdweq})), would have 
led to:
\be\label{wdweq-a}
\left[\frac{d^2}{da^2}-\frac{1}{a}\frac{d}{da}-U(a)\right]\psi=0.
\ee
This is just (\ref{WdWa}) with $\alpha=-1$, an ordering known to lead to exact solutions in the $a$ representation~\cite{vil0}. 

An important result follows from this realization. The two frameworks (Einstein-Cartan in the connection representation, and standard  quantum cosmology based on the metric representation and Einstein-Hilbert action) are Fourier duals {\it if there is no torsion and 
corresponding ordering is applied to both theories}. 
In view of (\ref{PBba}) the mapping is:
\bea\label{FT}
\psi_{a^2}(a^2)&=&\frac{3V_c}{l_P^2}\int \frac{db }{\sqrt{2\pi}} e^{-i\frac{3V_c}{\l_P^2}a^2 b}\psi_b(b).
\eea
A number of subtleties regarding the contour of integration explains why there is only one CS state, but
two possible solutions in the metric representation~\cite{CSHHV}.

For the purpose of this paper it is enough to stress that all equations to be examined in latter Sections are generalizations,
permitting a varying $\Lambda$, of the standard WdW equation expressed in the connection representation. 
We will prefer the $b$ representation here but, just as illustrated in this Section, all our solutions have duals  
in the $a$ representation according to a transformation of the form (\ref{FT}). We will study these complementary solutions
elsewhere.

\section{Parity-even branch on the quasi-Euler theory}\label{EulerPEth}
We now consider the simplest setting for a dynamical $\Lambda$: a theory with a quasi-Euler term only
($\gamma\rightarrow \infty$), in the branch where parity violations are not allowed ($c=0$, as in~\cite{alex1,alex2}). As explained in~\cite{MZ}, in this branch of the dynamics  the theory has two independent constraints, ${\cal H}$ and ${\cal V}$, because $N$ and $V$ are independent Lagrange multipliers. The two constraints form a closed algebra, so they are first class constraints with all the associated benefits. 

To briefly recap, the classical equations of motion for this system are:

\begin{align}
\frac{da}{dt} &= N\bigg(b-\frac{V\Lambda a}{3}\bigg) \\
\frac{db}{dt} &= N\frac{\Lambda a}{3} \\
\frac{d\Lambda}{dt} &= N\frac{V\Lambda^{2}}{6}\\
\frac{d\Pi}{dt} &= N\frac{2\Lambda^{3}a^{2}}{3}
\end{align}
We may access a gauge (with gauge transformations generated by the constraint ${\cal V}$) where $V=0$. In this gauge $\dot{\Lambda}=0$, which can be immediately solved to yield $\Lambda= \Lambda_{0}$ and so $\Lambda$ enters the remaining equations as an integration constant and the system reduces to General Relativity. 

Even though General Relativity in the presence of a cosmological constant is recovered classically, it is after integration of the classical equations of motion. It is conceivable that this equivalence between the theory and General Relativity does not persist at the quantum level.

We consider a representation diagonal
both in $b$ and $\Lambda$, so that $\psi=\psi(b,\Lambda)$. Since the wave-functions are diagonal in Lambda we can make them
dependent on any function of $\Lambda$. It will be notationally convenient to introduce the variable $\phi=1/\Lambda$, so that 
$\psi=\psi(b,\phi)$. 
In this representation, in view of commutators (\ref{com1}) and (\ref{com2}), the conjugate operators, therefore,
accept the representations:
\bea
\hat a^2&=&-\frac{i l_P^2}{3 V_c}\frac{\partial }{\partial b}\label{opr1}\\
\hat{ \Pi}& =& -\frac{2 i l_P^2}{3 V_c}\frac{\partial}{\partial \phi}\label{opr2}.
\eea
Thus, the two first class constraints are quantum mechanically represented by equations:
\bea
\hat{\cal H}\psi=\bigg(\frac{i \Lambda l_P^2}{9 V_c}\frac{\partial}{\partial b} + k+b^2 \bigg)\psi&=&0\label{Heq}\\
\hat{\cal V}\psi=\bigg(-\frac{i l_P^2}{9 V_c}\frac{\partial}{\partial \phi} -bk-\frac{1}{3}b^{3}\bigg)\psi&=&0.\label{Veq}
\eea
Notice that no ordering issues arise in the second equation (unlike in the first one, as already discussed). 
  
It is interesting that a straightforward generalization of the Chern-Simons state provides a solution to these
equations: 
\be\label{kodb}
\psi_{\tilde K}(b,\Lambda)={\cal N} \exp{\bigg(i\frac{ 9 V_c}{\Lambda l_P^2} \left(\frac{b^3}{3}+kb\right)\bigg)}.
\ee
This wave-function is just (\ref{kod0}) (in which $\Lambda$ is fixed parameter $\Lambda_0$) with the
constant $\Lambda_0$ replaced by its counterpart variable in this representation of dynamic operators $\hat {\frac{1}{\Lambda}}$ and $\hat \Pi$. In spite of its appearance, this is a non-trivial result. Notice also that $b$ now classically contains the torsion
(and so classically differs from $\dot a$ even on-shell)
 but we do not need to know that within the quantum mechanical theory.

Why does the Chern-Simons state generalize so easily? Given that the theme will recur throughout this paper, we should
illuminate this miracle from the start. With an eye on future generalization we write our constraint operators  
as:
\bea
\hat{\cal H}&=&\frac{i \Lambda l_P^2}{9 V_c}\frac{\partial}{\partial b} + h \label{Heqb}\\
\hat{\cal V} &=&-\frac{i l_P^2}{9 V_c}\frac{\partial}{\partial \phi} -\tau_{CS}\label{Veqb}
\eea
with functions $h$ and $\tau_{CS}$ taking the form:
\bea
h&=&b^2+k\\
\tau_{CS}&=&\frac{1}{3}b^{3} +bk
\eea
 for the theory in this Section.
To solve  $\hat {\cal H}\psi=\hat{\cal V}\psi=0$ we can try out ansatz:
\be\label{Sansatz}
\psi={\cal N} e^{\cal S}
\ee
resulting in
\bea
\frac{il_P^2}{9 V_c}\frac{\partial{\cal S}}{\partial b}&=& -h\phi\label{HJ1}\\
\frac{i l_P^2}{9 V_c}\frac{\partial {\cal S}}{\partial \phi}&=&-\tau_{CS}\label{HJ2}.
\eea
The integrability condition
\be\label{inteq}
\frac{\partial^2{\cal S}}{\partial b\partial  \phi }=\frac{\partial^2{\cal S}}{\partial  \phi\partial b}
\ee
is satisfied, since it is trivial to check that:
\be\label{int-cond}
\frac{\partial \tau_{CS}}{\partial b}=\frac{\partial (h\phi) }{\partial  \phi}=h.
\ee
So the ansatz (\ref{Sansatz}) does provide a solution, with:
\be\label{Ssol}
{\cal S}=i\frac{ 9 V_c}{\Lambda l_P^2} \tau_{CS},
\ee
which is the generalized CS state (\ref{kodb}). This works because the Chern-Simons functional and the
function $h$ appearing in the Hamiltonian constraint satisfy (\ref{int-cond}). 

We stress that this is all dependent on the choice of the factor $-\frac{3}{2}$ in front of the Euler term in (\ref{euler}).
Had we replaced it by $-\frac{3\theta }{2}$ (as in~\cite{alex2}), the factor $\theta$ would multiply $\tau_{CS}$ in (\ref{Veqb})
and in (\ref{HJ2}). For $\theta\neq 1$, no longer would the integrability condition be satisfied.

\section{The probability of  $\Lambda$ in the basic theory}\label{probL}

To extract physical implications of the solution (\ref{kodb}) the issue of the inner product is crucial. Regarding the CS state it
has even been claimed that a non-trivial inner product  is needed to remove the apparent
pathologies of the state. As pointed out in~\cite{randono1,randono2} this is not needed in the Euclidean theory,
where the standard inner product can be used to delta-normalize the CS state. The situation is the same for our adapted and generalized CS state, for the same reason (the state is a pure phase). As a more general matter, the issue of the probabilistic interpretation of ``cosmological'' wave-functions remains an open one \cite{Isham:1992ms}. Here we will look at the interpretation of (\ref{kodb}) in the context of three specific interpretations - the 
kinematical 
inner product, the conditional probability interpretation of the wavefunction, and the physical inner product for systems with first class constraints.

\subsection{The kinematical
inner product}
Given a solution $\psi(b,\phi)$, we may look to interpret $|\psi|^{2}$ as a probability density according to the following measure:

\be
d\mu_{b\phi}=2\left(\frac{3V_c}{l_P^2}\right)^2 db \, d\phi=
\frac{18V_c^2}{l_P^4} \frac{1}{\Lambda^2} db \, d\Lambda.
\ee
For the CS state (\ref{kodb}) one therefore predicts a uniform distribution in $\phi=1/\Lambda$ and $b$, and these variables are independent
random variables. Note, however, that in analogy with~\cite{CSHHV}, the specification of the domain of the variables is 
required to peg down the state. For example, in~\cite{CSHHV} it was found that, depending on the contour of the connection, the CS
wave-function can be the Fourier dual of either the Hartle-Hawking or the Vilenkin wave-functions.

Hence the distribution of $\Lambda$ is an inverse uniform distribution. If
$\phi$ is allowed to vary from  $-\infty$ to $\infty$ (so that the distribution is uniform in the ``distribution'' sense), the inverse distribution is a delta function:
\be\label{delta}
P(\Lambda)=\delta(\Lambda),
\ee
that is, we obtain a degenerate distribution, where $\Lambda$ is forced to be deterministically zero. The same would happen if
$\phi\in (0,\infty)$.

We can be more rigorous with the limiting procedure by introducing arbitrary IR and UV cut offs for Lambda and for $b$, otherwise $\int |\psi|^{2} d\mu_{b\phi}$ would diverge.
Let us assume first that $0<\Lambda_{IR}<\Lambda<\Lambda_{UV}$. Then, following standard results in probability,  the distribution of $\Lambda$ is a inverse uniform distribution in this domain, i.e. with probability distribution function:
\be
P(\Lambda)=\frac{\Lambda_{UV}\Lambda_{IR}}{\Lambda^2(\Lambda_{UV}-\Lambda_{IR})}\approx \frac{\Lambda_{IR}}{\Lambda^2}
\ee
where all we have assumed in the last approximation is that $\Lambda_{UV}\gg \Lambda_{IR}$. 
Clearly the integral of the distribution is dominated by the regions near the IR cut off (and indeed is divergent
if this is zero, in the limit leading to (\ref{delta}). 

We can also work out all the 
moments and cumulants of $\Lambda$, starting with:
\bea
\langle\Lambda\rangle_{S}&=&\frac{\Lambda_{UV}\Lambda_{IR}}{\Lambda_{UV}-\Lambda_{IR}}\
\ln\frac{\Lambda_{UV}}{\Lambda_{IR}}\approx \Lambda_{IR}\ln\frac{\Lambda_{UV}}{\Lambda_{IR}} \label{meanl}\\
\sigma_{S}^2(\Lambda)&=&\Lambda_{UV}\Lambda_{IR}-\langle\Lambda\rangle_{S}^2. \label{varl}
\eea
where $\langle\rangle_{S}$ denotes an expectation value according to the simple inner product. If we allow for $\Lambda<0$ (as we should) then similar cutoffs may be imposed to that branch.
Hence the average $\Lambda$ is dominated by the IR regulator. The sharpness of the peak depends on the 
UV regulator. If $\Lambda_{UV}$ is kept fixed as $\Lambda_{IR}\rightarrow 0$ we have effectively 
$P(\Lambda)=\delta(\Lambda)$.



\subsection{Conditional probability interpretation}
An alternative interpretation wavefunction $\Psi(b,\phi)$ is in terms of \emph{conditional probabilities} \cite{Page:1983uc}. If we consider some fixed value $b_{0}$, we can look to interpret the probability $P(\phi|b_{0})$ of being between $\phi$ and $\phi+d\phi$ given $b_{0}$ being as follows:

\begin{align}
P(\phi |b_{0})  & \propto  \frac{1}{\int |\psi(\phi,b_{0})|^{2} d\mu_{\phi}} |\psi(\phi,b_{0})|^{2}
\end{align}
where $d\mu_{\phi} = 2\bigg(3V_{c}/l^{2}_{P}\bigg)^{2}d\phi$. As in the case of the simplest inner product it is necessary to introduce IR and UV cutoffs for $\Lambda$ in order for $\int |\psi(\phi,b_{0})|^{2} d\mu_{\phi}$ not to diverge. Then for example we can introduce an expectation value for $\Lambda$ according to the measure $d\mu_{\phi}$ and conditional probability $P(\phi|b_{0})$:

\begin{align}
\langle\Lambda\rangle_{C} &=  \int \Lambda P(\phi|b_{0}) d\mu_{\phi}.
\end{align}
As $b_{0}$ is not present in $|\psi(\phi,b_{0})|^{2}$ and the measure $d\mu_{b\phi}$ is ``Cartesian'' in terms of $b$, therefore $\langle\rangle_{C} = \langle\rangle_{S}$ in this case.

\subsection{Physical inner product interpretation}
We can write the action (\ref{Faction2}) when $c=0$ in the more compact form, defining $Q^{A}= (b,\phi)$:

\begin{align}
\label{scomp}
S[Q,P,{\cal N}] = \int dt \bigg( P_{A}\dot{Q}^{A} - {\cal N}^{A}\bigg(P_{A}- \partial_{A}{\cal S}(Q)\bigg)\bigg) 
\end{align}
where ${\cal S} =  -18\kappa V_{c}\bigg(\frac{b^{3}}{3}+bk\bigg)\phi$.
Then it can be checked that the (first class) constraints ${\cal H}=0$ and ${\cal V}=0$ are of the form:

\begin{align}
P_{A} &= \frac{\partial {\cal S}}{\partial Q^{A}} .
\end{align}
(This is closely related to integrability condition (\ref{inteq}) being satisfied.)
An approach to the interpretation of quantized theories possessing first class constraints of this form is to define a \emph{physical} inner product as follows \cite{Henneaux:1992ig}

\begin{align}
\label{physprod}
(f|g) &\propto \int \prod_{A'} dQ^{A'}\prod_{A}\delta(\chi_{A})|\det[\chi_{A},G_{B}]|f^{*}g
\end{align}
where $G_{A}$ denote the constraints $G_{A} = P_{A} - \frac{\partial {\cal S}}{\partial Q^{A}} = 0$, whilst the $\chi_{A}$ should be good gauge fixing conditions, here $\chi_{A} = a_{AB}(Q)(Q^{B}-f^{B})$ where $f^{B}$ does not depend on $Q^{A}$ and $\det a_{AB} \neq 0$, which guarantees that $(f|g)$ is invariant under changes of the gauge conditions. 

A motivation for the introduction of (\ref{physprod}) is to avoid unphysical divergences of the inner product caused by integration over unbounded ``gauge'' degrees of freedom. Note that in this case we have as many first class constraints as we do coordinates $Q^{A}$, and so there are an equal number of terms in each product in (\ref{physprod}). 

By way of example, classically the constraint ${\cal V}$ generates shifts in $\phi$ ($\{\phi,{\cal V}\}=1/6$) and so for a given $\phi(t)$ we may find a gauge where $\dot{\phi}=0$. From the $\phi$ equation of motion $\dot{\phi}=V/6$ this implies that $V=0$. Furthermore we may adopt the `proper time' spacetime gauge $N=1$; in this gauge the Euler-Lagrange equations imply that $\dot{\phi}=0$, with solution $\phi -\phi_{0}=0$, for constant $\phi_{0}$, whilst the remaining equations reduce to those of general relativity with cosmological constant $\Lambda_{0} = 1/\phi_{0}$; consequently,  we then have a gauge condition $b = \dot{a}_{GR(k)}(t)$, where $\dot{a}_{GR(k)}$ is the classical time derivative of the scale factor for de Sitter space with slicing dictated by the curvature $k$.

This implies the following form for the physical inner product $(f|g)$:

\begin{align}
(f|g) &\propto \int d\phi db \delta(\phi-\phi_{0})\delta(b-\dot{a}_{GR(k)}(t)) f^{*} g.
\end{align}
Using the solution (\ref{kodb}) then we see that

\begin{align}
(\psi_{\tilde{K}}|{\cal F}(\Lambda,b) |\psi_{\tilde{K}}) \propto {\cal F}(\Lambda_{0},\dot{a}_{GR(k)}(t)) \label{flb}.
\end{align}
If the constant of proportionality is chosen such that $(\psi_{\tilde{K}}|\psi_{\tilde{K}})=1$ then the proportionality in (\ref{flb}) becomes an equality and we see for example that $(\psi_{\tilde{K}}|\Lambda^{n}|\psi_{\tilde{K}}) =\Lambda^{n}_{0}$ and hence all moments of $\Lambda$ are determined by the classical value adopted in the gauge fixing choice.

\section{Parity-odd branch of Euler theory}\label{EulerPOth}
As explained in~\cite{MZ} the dynamics is radically  different in the parity-odd branch of the theory, 
where $c=0$ is not imposed, that is, if we switch on Cartan's spiral staircase~\cite{spiral}. 
Then, the action takes the form (\ref{Faction2}) for a general $c$. Variation with respect to $c$ leads to the
equation of motion:
\begin{align}
(2Na  - bV) c = 0 \label{ceq}.
\end{align}
This is trivially satisfied if $c=0$ is imposed, but otherwise it implies that the two Lagrange multipliers are 
related by $V = 2N a/b$. Omitting other constraints that may be present in this sector, we focus on the WdW equation which can be written as follows,
\be\label{WdWspiral}
\left(\hat {\cal H}+\frac{1}{b}\hat {\cal V}\right)\psi=0.
\ee
We first seek a particular solution by investigating whether
$\hat {\cal H}\psi=\hat {\cal V}\psi=0$ is possible. This amounts to solving the same problem as in Section~\ref{EulerPEth}
with the functions $h$ and $\tau_{CS}$ defined in (\ref{Heqb}) and (\ref{Veqb}) replaced by:
\bea
h&=&b^2+k-c^2\label{hc1}\\
\tau_{CS}&=&\frac{1}{3}b^{3} +b(k-c^2).\label{tCSc1}
\eea
Here $c$ can be thought of as a number labelling classes of distinct solutions to (\ref{WdWspiral}). Once again, an ansatz of the form (\ref{Sansatz}) leads to a solution, since the integrability condition (\ref{inteq}) for the resulting equations (\ref{HJ1}) and (\ref{HJ2}) 
is still satisfied: Eq.~(\ref{int-cond}) still works for the $h$ and $\tau_{CS}$ given by (\ref{hc1}) and (\ref{tCSc1}). The solution, therefore, is still (\ref{Ssol}), with $\tau_{CS}$ updated to (\ref{tCSc1}), that is:
\be\label{kodbc1}
\psi_{\tilde K}(b,\Lambda)= {\cal N} \exp{\bigg(i\frac{ 9 V_c}{\Lambda l_P^2} \left(\frac{b^3}{3}+(k-c^2) b\right)\bigg)}.
\ee
As detailed in Appendix~\ref{CSapp}, this is once more the CS wave-function in the context of this theory and reduction. 

However, this is not the most general solution because $\hat {\cal H}\psi$ can now be non-zero if balanced by a suitable $\hat{\cal V}\psi$ in (\ref{WdWspiral}).
 In order to find the most general solution we try out an ansatz where the normalization constant ${\cal N}$ in (\ref{kodbc1}) is replaced by a generic amplitude:
\be\label{kodbc2}
\psi_{\tilde K}(b,\Lambda)= {\cal A}(b,\Lambda,c) \exp{\bigg(i\frac{ 9 V_c}{\Lambda l_P^2} \left(\frac{b^3}{3}+(k-c^2) b\right)\bigg)}.\nn
\ee
Insertion in~(\ref{WdWspiral}) leads to:
\be\label{ampeq0}
\frac{1}{\phi} \partial_b{\cal A}=\frac{1}{b}\partial_{\phi}{\cal A}
\ee
so that:
\be\label{ampeq1}
{\cal A}={\cal A}(\phi b,c)={\cal A}\left(\frac{b}{\Lambda},c\right).
\ee

We now present the resolution of an apparent paradox. Inspection of the single constraint (\ref{WdWspiral}) reveals that $k$ and $c^2$ drop out:
\be\label{comboconstr}
\left(\frac{2}{3}b^2 -\frac{\Lambda}{3}\hat a^{2} + \frac{\hat \Pi}{6b} \right)\psi=0.
\ee
This equation can be solved directly, with a particular solution:
\be\label{kodbc3}
\psi_{\tilde K}(b,\Lambda)= {\cal N}' \exp{\bigg(i\frac{ 9 V_c}{\Lambda l_P^2} \left(\frac{b^3}{3} \right)\bigg)}
\ee
where neither $c$ or $k$ feature. 
This seems in contradiction with (\ref{kodbc1}). However this is not the case, since  (\ref{kodbc1}) and  (\ref{kodbc3}) are only particular solutions. The most general solution to (\ref{comboconstr}) takes the form:
\be\label{kodbc4}
\psi_{\tilde K}(b,\Lambda)= {\cal B}\left(\frac{b}{\Lambda},c\right) \exp{\bigg(i\frac{ 9 V_c}{\Lambda l_P^2} \left(\frac{b^3}{3}\right)\bigg)}
\ee
and this is equivalent to: 
\be\label{kodbc2}
\psi_{\tilde K}(b,\Lambda)= {\cal A}\left(\frac{b}{\Lambda},c\right) \exp{\bigg(i\frac{ 9 V_c}{\Lambda l_P^2} \left(\frac{b^3}{3}+(k-c^2) b\right)\bigg)}
\ee
if:
\be
{\cal A}={\cal B}
\exp{\bigg(-i\frac{ 9 V_c}{\Lambda l_P^2} (k-c^2) b \bigg)}.
\ee
The fact that the phase relating the two is a function of $b/\Lambda$ (and of $c$) ensures consistency.

It is interesting to note that the amplitude, and therefore the probability, is a function of the classical constants of motion.
Indeed it can be proved~\cite{MZ} that for the pure Euler case all solutions with $c\neq 0$ satisfy:
\bea
c&=&c_0\\
b&=&\beta_0 \Lambda.
\eea
The quantum theory then is a theory of the probabilities for $c_0$ and $\beta_0$. 
Unfortunately, the WdW equation by itself does not fix this amplitude. Not only do we encounter the same potential
ambiguities in the inner product as in the parity invariant case (whose relevance is debatable), but now the quantum
theory fails to uniquely specify the distribution according to each of the inner products considered in Section \ref{probL}. Note that as there is now only one first-class constraint, the form of the physical inner product is altered, retaining a non-trivial integration over ${\cal B}^{*}{\cal B}$ over the degree of freedom that remains after gauge fixing.


\section{The Pontryagin case}\label{PontTH}
The situation is dramatically different if we allow for a Pontryagin term. 
In this case there is no parity-even branch in MSS, and the dynamics is necessarily subject to a
single constraint, rather than two. Below we spell out the classical dynamics (only partly studied in~\cite{MZ}). 
We then study the resulting WdW equation. We find that it no longer accepts the CS 
wave function as a particular solution, because $\hat{\cal H}\psi=0$ is now inconsistent with
$\hat{\cal V}\psi=0$. Nonetheless, we are able to find the general solutions of the relevant equation 
which combines the two constraints into the only one applicable to this theory.

\subsection{Dirac analysis of the classical theory}\label{PontHam}
If we add a Pontryagin term to our action,  we can still cast the FRW-reduced action into form (\ref{Faction2}), i.e. same conjugate variables and Poisson brackets, and two constraints, ${\cal H}$ and ${\cal V}$. To to this we should start from Eq.147 in~\cite{MZ}, which we reproduce here
for convenience (with the same factors restored, as in (\ref{Faction2})):
\begin{align}
\label{SFRW1}
S^{g}
&=3\kappa V_c \int dt \bigg( \bigg(2(k-c^2+b^2)Na + 2\dot{b}a^{2}\bigg)-\frac{2\Lambda}{3} Na^{3}  \nn\\
&-\frac{6}{\Lambda}(k-c^2+b^2)\bigg(\dot{b}-\frac{1}{\gamma}\dot{c}\bigg)+\frac{12}{\Lambda}bc\bigg(\dot{c}+\frac{1}{\gamma}\dot{b}\bigg)\bigg).
\end{align}
After the same algebraic manipulations which lead to (\ref{Faction2}) for the quasi-Euler theory, we obtain an expression identical in everything to  (\ref{Faction2}) except that the constraint ${\cal V}$ is replaced by:
\be
{\cal V}=\frac{\Pi}{6} - \left(\frac{b^{3} }{3}
+b(k-c^2) \right)
-\frac{1}{\gamma}\left(\frac{c^3}{3} - c(b^2+k)\right).
\ee
Variation of the action with respect to $c$ now imposes the equation upon $N$ and $V$:
\begin{align}
V(b^{2}-c^{2}+k+2bc\gamma) &= 4ca \gamma N
\end{align}
as already noted in~\cite{MZ}. 

We see that no longer does a $c=0$ branch exist. Instead, the $V$ and $N$ must always be 
related, so that the two constraints reduce to  a single constraint of the form:
\be\label{PontConst}
\left(b+\frac{b^2+k-c^2}{2\gamma c}\right){\cal H}+{\cal V}\approx 0.
\ee
This completes the classical analysis of the quasi-Euler-Pontryagin  theory in MSS.

\subsection{Solving the WdW equation}\label{PontWdW}
Quantization in the representation diagonal in $b$, $\Lambda$ and $c$ therefore proceeds according to the representations (\ref{opr1}) and (\ref{opr2}) applied to (\ref{PontConst}). 
Once again, the operators $\hat {\cal H}$ and $\hat {\cal V}$ can be cast in the form (\ref{Heqb}) and (\ref{Veqb}), with
\bea
h&=&b^2+k-c^2\label{hPont}\\
\tau_{CS}&=&\frac{1}{3}b^{3} +b(k-c^2)+\frac{1}{\gamma}\left(\frac{c^3}{3}-c(b^2+k)\right).\label{CSPont}
\eea
We want to solve:
\be\label{wdwpont}
\left[\left(b+\frac{h}{2\gamma c}\right)\hat {\cal H}+\hat{\cal V}\right]\psi=0
\ee
As before, one particular solution could result from trying out $\hat {\cal H}\psi=\hat  {\cal V}\psi=0$.
The generalized CS state, solving $\hat  {\cal V}\psi=0$ would take the form:
\be\label{kodPONT}
\psi(b,\Lambda)= {\cal N} \exp{\bigg(i\frac{ 9 V_c}{\Lambda l_P^2} \tau_{CS}\bigg)}
\ee
It can be checked (see Appendix~\ref{CSapp})  that this is still the CS functional applied to FRW
once a complex coupling constant is introduced to account for the Pontryagin term. We refer the reader to the 
Appendix for more details. In Appendix~\ref{CSapp} it is also explained why in a general setting 
$\hat {\cal H}\psi=0$ is incompatible with $\hat  {\cal V}\psi=0$ if a Pontryagin term is present. 
Within MSS this can be understood revisiting the argument at the end of Section~\ref{EulerPEth}, explaining the 
``miraculous'' generalization of the CS state to the Euler theory.
Recall that
$\hat{\cal H}\psi=\hat{\cal V} \psi=0$, implies the integrability condition (\ref{inteq}), amounting to:
\be
\partial_b \tau_{CS}=h.
\ee
As an examination of (\ref{hPont}) and (\ref{CSPont}) shows, this 
is no longer valid if $\gamma$ is finite. Therefore no longer is an extension of the CS state a solution to 
this theory.

Trying out the obvious modification of the CS state 
we set:
\be
\psi(b,\Lambda)= {\cal A}(b,\Lambda,c) \exp{\bigg(i\frac{ 9 V_c}{\Lambda l_P^2} \tau_{CS}\bigg)}.
\ee
Inserting into (\ref{wdwpont}) leads to:
\be
\left(b+\frac{h}{2\gamma c}\right)\left(\frac{i \Lambda l_P^2}{9 V_c}\frac{\partial{\cal A}}{\partial b}
-\frac{2bc}{\gamma}{\cal A}\right)-\frac{i l_P^2}{9 V_c}\frac{\partial {\cal A}}{\partial \Lambda^{-1}} =0.
\ee
It can readily be checked that if:
\be
{\cal A}={\cal A}\left(\frac{c^2}{\Lambda}\left(b+\frac{h}{2\gamma c}\right)\right)
\ee
then this becomes the ODE:
\be
\frac{i \Lambda l_P^2}{9 V_c}{\cal A}'=-2{\cal A}.
\ee
After the obvious integration we therefore find the {\it particular} solution:

\begin{align}
\label{kodbcg1}
\psi_{\tilde K}(b,\Lambda)&={\cal N} \exp{\bigg(i\frac{ 9 V_c}{\Lambda l_P^2} \left(\tau_{CS} +2c^2\left(b+\frac{h}{2\gamma c}\right)\right) \bigg)}\nn\\
&={\cal N} \exp{\bigg(i\frac{ 9 V_c}{\Lambda l_P^2} \left(\frac{b^3}{3}+b(k+c^2)-\frac{2c^3}{3\gamma}\right)\bigg)}
\end{align}
(where the $+$ sign in $c^2$ is NOT a typo). Hence a wave-function 
of the form $\psi={\cal N} e^{\cal S}$ is still possible, but not with $S\propto \tau_{CS}$.

However, as for the Euler theory, (\ref{kodbcg1}) is only a particular solution, not the most general one. One can 
further try out an ansatz of the form (\ref{kodbcg1}) but with the constant ${\cal N}$ replaced by a generic amplitude ${\cal A}(b,\Lambda,c)$.
This leads to equation:
\be
\left(b+\frac{h}{2\gamma c}\right) \Lambda \partial_b{\cal A}=\partial_{\Lambda^{-1}}{\cal A}
\ee
which is a generalization of (\ref{ampeq1}). By analogy, we can try 
\be
{\cal A}={\cal A}\left(\frac{f(b)}{\Lambda}\right)
\ee
leading to:
\be
\left(b+\frac{h}{2\gamma c}\right) f' =f
\ee
which integrates to:
\be
f(b)=\exp{\left(2\gamma c\frac{\arctan\frac{b+\gamma c}{\sqrt{k- c^2 (1+\gamma )}} }{\sqrt{k-c^2(1+\gamma)} }\right)}.
\ee
For solutions with $c^2(1+\gamma)\ge k $ this becomes:
\be
f(b)=\exp {\left(\frac{\gamma c}{ \sqrt{k-c^2(1+\gamma)}}\ln\frac{1- 
\frac{b+\gamma c}{\sqrt{k- c^2 (1+\gamma )}}}
{1+
\frac{b+\gamma c}{\sqrt{k- c^2 (1+\gamma )}}}\right)},
\ee
or finally:
\be
{\cal A}={\cal A}\left(\phi\left(\frac{1- 
\frac{b+\gamma c}{\sqrt{k- c^2 (1+\gamma )}}}
{1+
\frac{b+\gamma c}{\sqrt{k- c^2 (1+\gamma )}}}\right)^{\frac{\gamma c}{ \sqrt{k-c^2(1+\gamma)}}}\right).
\ee

As in the pure Euler case, if parity is broken, the quantum theory fails to uniquely specify the distribution according to each of the inner products considered in Section V. We defer to a future investigation a more complete study of the Pontryagin solution, and the meaning
of the predicted amplitude functional dependence. Qualitative differences with the purely Euler case do exist. 

\section{Comparison to other models of dynamical $\Lambda$}
\label{comparison}
We have seen that for the case $\gamma\rightarrow \infty$ and $c=0$, we have a model which in FRW symmetry possesses a symmetry allowing one to find a `conformal gauge' where $\dot{\Lambda}=0$ via the classical equations of motion. This equation of motion - the dynamical recovery of a cosmological constant\footnote{See also \cite{Jirousek:2018ago} and \cite{Hammer:2020dqp} for interesting recent approaches.} also appears if one adds to general relativity an action for a gauge field 3-form $A = \frac{1}{3!}A_{\mu\nu\rho}dx^{\mu}dx^{\nu}dx^{\nu}$ and Lagrangian $-F^{\mu\nu\alpha\beta}F_{\mu\nu\alpha\beta}\sqrt{-g}$ where $F=dA$ \cite{weinberg}. A model which also produces this equation is the so-called covariant unimodular gravity \cite{Henneaux:1989zc} where one considers adding to the action of General Relativity an action

\begin{align}
S_{UG}[\Lambda,T^{\mu}] &=  -\frac{1}{8\pi G}\int \sqrt{-g} \bigg(\Lambda + T^{\mu}\partial_{\mu}\Lambda\bigg) \label{umog}
\end{align}
In FRW symmetry, we can take $T^{\mu} = T(t)\delta^{\mu}_{t}$ and hence a contribution to the MSS Lagrangian the momentum of $\Lambda$ can be thought of as due to the genuinely new degree of freedom $T$ as opposed to the momentum $\Pi$ of $\Lambda$ in the case (\ref{Faction2}) which is instead related to the Chern-Simons time $\tau_{CS}$ which is built from the spin connection itself. This turns out to be a key difference. In recent work \cite{Gryb:2018zoz,Gryb:2018whn,Gryb:2018kxe,Gielen:2020abd}, the quantization of General Relativity coupled to a massless scalar field $\xi$ and the action (\ref{umog}) was considered. Considering $T$ instead as a configuration variable, then $\Lambda$ can be associated with its momentum $P_{T}$; the appearance of $\Lambda$ in the Hamiltonian constraint then leads to $P_{T} \rightarrow -i\hbar \partial_{T}$ appearing linearly in the Hamiltonian constraint, resulting in a time-dependent Schrodinger equation in terms of $T$ with solutions $\psi(T,a,\xi)$ such that an inner product $\langle \psi_{1}| \psi_{2} \rangle_{T}$ is preserved by evolution with respect to $T$.
Intriguingly, in the quantum theory, the big-bang singularity that would classically result from General Relativity coupled to a massless scalar field is avoided, with the expectation value of the scale factor $\langle \psi |a(T)|\psi \rangle_{T}$ remaining non-zero for all values of $T$. Here, quantum states consist of a superposition of values of $\Lambda$, which for example may be chosen to take the form of a Gaussian `wavepacket' peaked at some value $\Lambda_{c}$ - this in contrast to the expectation value (\ref{meanl}) which takes a value related to putative infrared and ultraviolet cutoffs that might be assumed for the configuration variable $\Lambda$.

\section{Matter Couplings}
\label{matter}

It is natural to wonder to what extent coupling to matter will complicate the previous results. Towards this end, we briefly consider the introduction of a simple matter component: a massless scalar field $\xi$. It can be shown that the reduced action in the presence of this scalar field can be written as: 

\begin{align}
\label{Faction3}
S&=3\kappa V_c \int dt \bigg(  2a^{2} \dot{b}+P_{\phi} \dot{\phi}+ P_{\xi}\dot{\xi}+ 2Na {\cal H}+V{\cal V}\bigg)\\
{\cal H} &= b^{2}+k-c^{2}-\frac{\Lambda}{3}a^{2} - \frac{1}{4a^{4}}P_{\xi}^{2}\\
{\cal V} &= \frac{P_{\phi}}{6}-\bigg(\frac{b^{3}}{3}+b\bigg(k-c^{2}\bigg)\bigg).
\end{align}
Suppose we proceed as before and vary $c$, yielding an equation that still either allows us to choose $c=0$ or $c\neq 0$. If we allow for $c\neq 0$ the two Lagrange multiplies $N$ and $V$ are fixed in terms of one another. Introducing the canonical momentum of $b$, $P_{b} = 2a^{2}$ we get the following, single, Hamiltonian constraint term:

\begin{align}
{\cal N}{\cal H}_{b} &= {\cal N}\bigg(\frac{2b^{2}}{3}-\frac{P_{\xi}^{2}}{P_{b}^{2}}+\frac{1}{6}\bigg(\frac{P_{\phi}}{b}-\frac{P_{b}}{\phi}\bigg)\bigg) \label{hamb}
\end{align}
where ${\cal N} \equiv 2aN$. The inverse power of $P_{b}$ present in (\ref{hamb}) poses a challenge in terms of quantization. One may look to rescale ${\cal N} \rightarrow {\cal N}/P_{b}^{2}$ to remove this term at the cost of creating a term $b^{2}P_{b}^{2}$ and the introduction of new operator ordering ambiguity.

Alternatively one may use $a$ as the configuration variable with canonical momentum $P_{a}=-4a b$. This instead results in the following, single, Hamiltonian constraint term:

\begin{align}
{\cal N}{\cal H}_{a} &= {\cal N}\bigg(\frac{P_{a}^{2}}{24a^{2}}-\frac{P_{\xi}^{2}}{4a^{4}}-\frac{2a P_{\phi}}{3P_{a}}-\frac{a^{2}}{3\phi}\bigg)
\end{align}
We see that now there is a term inversely proportional to $P_{a}$ in the Hamiltonian constraint. Again one may look to redefine ${\cal N}$, in this case ${\cal N}\rightarrow {\cal N}/P_{a}$ at the expense of introducing terms such as $a^{2}P_{a}$ in the Hamiltonian constraint and new ordering ambiguity/ The fact that it is $P_{\xi}$ that results in an awkward term in the connection representation and $P_{\phi}$ that results in an awkward term in the metric representation may suggest that the structure of the field $\phi$ sits best alongside a perspective of $b$ being the basic gravitational variable, whilst $\xi$ seems more easier to consider from the perspective of $a$ being the basic gravitational variable.

\section{Conclusions}

The main purpose of this paper was to derive the quantum cosmology of an extension of the Einstein-Cartan theory where the 
cosmological constant $\Lambda$ can be directly promoted to a variable, without conflict with the Bianchi identities, 
due to a balancing torsion. This theory is a quasi-topological theory where Euler and Pontryagin invariants have their topological nature disrupted by a non-constant $\Lambda$. We found that for the Euler-based theory the solutions in the connection and $\Lambda$ representation are straightforward generalization and adaptations of the Chern-Simons (or Kodama) state. That is, they take exactly the same form as the CS state, but with parameter $\Lambda_0$ replaced by variable $\Lambda$, and the CS functional evaluated for  the appropriate expression for the connection. This fact is true for both the parity-even and parity-odd branches of the theory, with a small adaptation. For the parity-odd branch, where the CS functional has an imaginary and real parts, only the imaginary part of the functional, multiplied by $i$, appears in the solution. Hence, the solution always has an imaginary exponent. For the Euler based theory, the CS state can also be multiplied by a non-constant amplitude, although the amplitude may equally well be constant.

This is an interesting result. It can be shown that well-known solutions in the metric representation for standard Einstein theory, such as the Hartle-Hawking and Vilenkin wave-functions, already are the Fourier transform of the CS state reduced to MSS. Specifically, they are the same quantum state in complementary representations (see~\cite{CSHHV} for a full qualification of this statement). We now see that straightforward generalizations and adaptations of the CS state can accommodate a variable Lambda in quasi-topological theories based on the Euler invariant. We explained this 
 non-trivial structure 
in terms of the consistency of the Hamiltonian and conformal constraint, both solved by the generalized and adapted CS state. Unfortunately this  structure  does not survive in theories based on the Pontryagin 
(which only have a parity-odd branch) because for these the two constraints become incompatible. Nonetheless we were able to derive the general solutions for these as well. 

What can we conclude from this? The CS state solves the full, non-perturbative Hamiltonian constraint in the self-dual, or Ashtekar formulation~\cite{jackiw,witten,kodama,lee1,lee2}. Nonetheless, the state has a poor reputation, at least in its Lorentzian formulation. 
A number of fair criticisms have been voiced (e.g.~\cite{witten}), namely regarding its non-normalizability, 
CPT violating properties (and consequent impossibility of a positive energy property), and lack of gauge invariance under large 
gauge transformations. All of these pathologies result  from the fact that the state's phase is not purely imaginary. Hence they do not apply to the adapted CS state derived here, in which the phase is proportional to $i\Im Y_{CS}$, and so purely imaginary and only sensitive to the imaginary part of $Y_{CS}$. Then our Lorentzian solutions resemble the solutions for the Euclidean theory, for which the pathologies listed above are not present~\cite{randono1,randono2}. For example, 
if the exponential has a pure phase, then delta function normalization is possible, as in standard quantum mechanics, with the standard inner product, for plane waves. 

Why have we bypassed so many problems with the CS state? In part this is because we are working in MSS, but also because we have worked with an explicitly real theory, as explained in Appendix~\ref{CSapp}. Obviously issues of interpretation still depend on the inner product chosen, but one does not longer need to invoke non-trivial inner products as a way to save the CS state.
It remains to be seen whether this structure survives the full non-perturbative theory. That is the purpose of~\cite{MSZ1}, for which this paper provides the blueprint. 

In future work we will also examine the metric duals of the wave-functions derived in this paper, that is the equivalents of the Hartle-Hawking and Vilenkin wave-functions for these theories. In the process we hope to shed further light on the predictions made for the cosmological constant in these models, and issues such as ``creation out of nothing''. 

We close with a speculation. Could there be a relation between our work and that of~\cite{Steph-calc},
where quantum gravity is seen as a Fermi liquid?
In that work the CS state is also trivially generalized for a dynamical Lambda. A further point of analogy 
is parity violation. However, closer inspection reveals that the technical overlap is actually minimal. 
First, 
in our work $\Lambda$ becomes a truly independent dynamical variable in phase space (with a conjugate momentum forced
to be related to the CS functional by a constraint), whereas in~\cite{Steph-calc} it is a function of the Ashtekar connection.
Then, the work in~\cite{Steph-calc} relies on the interesting degenerate solutions proposed by Jacobson~\cite{jacobson}:
it is hard to see how they would fit in our MSS reduction. Finally, we do not need to invoke fermions, directly or effectively, 
at any stage. 
Rather than direct overlap, we therefore feel that a valuable synergy between our paper and~\cite{Steph-calc}
might be obtained by {\it combining} the two approaches.  
Could there be a topologically non-trivial sector with the same origin as that in~\cite{Steph-calc} for the theory proposed 
here? Would this lead to a  formula/distribution for Lambda more natural than the one obtained in either approach? 
Could the setting proposed here provide a more natural realization of the radical proposal 
that quantum gravity is a Fermi liquid?

\section*{Acknowledgements}
We thank S. Alexander,  D. Jennings and A. Vilenkin for helpful comments. JM was funded by the STFC Consolidated
Grant ST/L00044X/1.
TZ was supported by the funds from the European Regional Development Fund and the Czech Ministry of Education, Youth and Sports (M\v{S}MT): Project
CoGraDS - CZ.02.1.01/0.0/0.0/15\_003/0000437. 

\appendix

\section{Relation to the Chern-Simons state}\label{CSapp}

In this Appendix we show that the various wave-functions found throughout this paper are MSS reductions of a simple
adaptation of the CS state, valid for all cases considered. We first present it in its full form, then reduce it to MSS.

\subsection{General form of the adapted state}
The adaptation we use applies to explicitly real theories, and can be used either in the quasi-topological theory 
or in the standard Palatini framework. It will be studied elsewhere in more detail (see also~\cite{randono1,randono2,wieland}). Here we present the main results.

Starting with the pure Palatini case, the adapted CS wave-function is:
\bea\label{kodrealth}
\psi(A)&=&{\cal N}\exp {\left(-\frac{3i}{2l_P^2\Lambda}\Im Y_{CS}\right)}
\eea
where 
\be
Y_{CS}=\int {\cal L}_{CS}=\int  A^i dA^i +\frac{2}{3}\epsilon_{ijk} A^i A^j A^k\
\ee
is the Chern-Simons functional. Here $i,j,k$ are the $SU(2)$ indices of Ashtekar's self-dual connection:
\be\label{ashcon}
A^i=\omega^i + i  \omega^{0i}
\ee
with $\omega^i=-\frac{1}{2}\epsilon^i_{jk}\omega^{jk}$. We stress the distinction between the 
Immirzi parameter used to define (\ref{ashcon}) and $\gamma$ used to parameterize the Pontryagin term
(and also the parameter in the factor of the Holst/Nieh-Yan term, if used). They can all be different, and in particular
we may use a connection complex variable whilst keeping the action real, as is the case here.

The wave-function (\ref{kodrealth}) solves the Hamiltonian constraint for the Palatini theory (i.e., with fixed $\Lambda$). 
Although this is not needed for the purpose of a MSS reduction,  a simple adaptation also solves the Hamiltonian constraint of
the quasi-topological theory, where $\Lambda$ becomes dynamical. In particular, if $\Lambda$ varies in space (beyond MSS) the solution generalizes to:
\bea\label{kodrealtha}
\psi(A)&=&{\cal N}\exp {\left(-\frac{3i}{2l_P^2}\Im Y_{\Lambda}\right)}
\eea
where 
\be
Y_{\Lambda}=\int \frac{{\cal L}_{CS}}{\Lambda}
\ee
i.e. it is enough to shift $\Lambda$ in (\ref{kodrealth}) inside the integral $Y_{CS}$.

The $\cal V$ constraint leads to the same state {\it if there is no Pontryagin term}, as we now sketch (see~\cite{MSZ1} for details). The quasi-topological terms
can be written in terms of the Ashtekar connection~\cite{alex0} as:
\be\label{SnewSD}
S_{new}=S_{Eul}+ S_{Pont}=-\frac{\kappa}{2}\int \frac{\zeta}{\Lambda} R^i R^i +\frac{\zeta^\star}{\Lambda} \bar R^i \bar R^i.
\ee
Crucially, we allow the coupling parameter $\zeta$ to be complex, and take the value\footnote{If we want to depart from the miracle factor $-3/2$ in the Euler term,  as in~\cite{alex2}), we should set $\zeta=3\left(\frac{1}{\gamma} +i\theta\right)$ and explore $\theta\neq1$.}:
\be\label{zeta}
\zeta=3\left(\frac{1}{\gamma} +i\right).
\ee
Indeed, 
appealing to the standard formulae in the self-dual formalism: 
\be
R^i R^i=\frac{1}{2}{\left( R^{AB}R_{AB}  -\frac{i}{2}
 \epsilon_{ABCD}R^{AB}R^{CD}\right)}
\ee
we see that  (\ref{SnewSD}) does follow from (\ref{Sg}).  
But given that:
\be\label{potential}
R^i R^i=d
{\cal L}_{CS}=d\left(A^idA^i+\frac{2}{3}\epsilon_{ijk} A^i A^j A^k\right)
\ee
we have:
\be
S_{new}=-\kappa \int \frac{1}{\Lambda}d\Re( \zeta {\cal L}_{CS}).
\ee
After a 3+1 split and an integration by parts mimicking the one done in MSS (and after resolving a 
subtlety for when $\Lambda$ varies in space~\cite{MSZ1}) we find that 
the conjugate momentum to $\phi=1/\Lambda$ defined by:
\be
\{\phi({\bf x}),\Pi({\bf y})\}=\frac{1}{\kappa}\delta({\bf x}-{\bf y})
\ee
is subject to constraint:
\be
{\cal V}=\Pi - \Re( \zeta {\cal L}_{CS})\approx 0
\ee
which now must be enforced point by point. Therefore, quantum mechanically:
\be
\hat \Pi({\bf x}) =-2il_P^2\frac{\delta}{\delta\phi({\bf x})}
\ee
so that  $\hat {\cal V}\psi=0$ is solved (point by point) by:
\bea\label{realKod}
\psi&=&{\cal N}\exp{\left(\frac{i}{2 l_P^2 }\Re \zeta Y_{\Lambda }\right)}
\eea
where again all we need to do is shift the factor of $1/\Lambda$ inside the integral. 
Using (\ref{zeta}), this is equivalent to:
\bea\label{realKodb}
\psi
&=&{\cal N}\exp{\left(\frac{i}{2 l_P^2}
(-3\Im Y_{\Lambda} +\frac{3}{\gamma}\Re Y_{\Lambda})\right)}\label{kodrealthB}.
\eea 
This is the CS required by the ${\cal V}$ constraint. 
By comparing (\ref{kodrealth}) with this expression, we rediscover, in more generality, the result found in Section~\ref{PontTH}  that for the Pontryagin theory the two constraints are incompatible. 

\subsection{Reduction to MSS}
We can now verify that various solutions found throughout this paper are the MSS reductions of wave-function
(\ref{realKodb}).
We evaluate $A^I$ by plugging the results of~\cite{MZ} (see Appendix I) into the definition (\ref{ashcon}).
For $k=0$ we have:
\be
A^i=(c+ib)E^i,
\ee
so that:
\be
Y_{CS}=\int \frac{2}{3}(c+ib)^3\epsilon_{ijk}E^iE^jE^k=4 V_c (c+ib)^3.
\ee
The calculation for $k\neq 0$ and leads to:
\be
Y_{CS}=4 V_c (c+ib)((c+ib)^2 - 3k).
\ee
Given that:
\bea
\Re Y_{CS}&=&4 V_c (c^3-3c(b^2+k))\\
\Im Y_{CS}&=&-4V_c(b^3+3b(k-c^2)),
\eea
Inserting into (\ref{kodrealthB}) we finally find:
\begin{widetext}
\be
\psi_K= {\cal N} \exp{\bigg[i\frac{ 9 V_c}{\Lambda l_P^2} \left(
\frac{1}{3}b^{3} +b(k-c^2)+\frac{1}{\gamma}\left(\frac{c^3}{3}-c(b^2+k)\right)
\right)\bigg]}.\label{kodrealMSS}
\ee
\end{widetext}
We see that indeed this corresponds to all the solutions for MSS found throughout this paper
(i.e., Eqns.~(\ref{kod0}), (\ref{kodb}), 
 (\ref{kodbc1}) and (\ref{kodPONT})). This is true even for the cases where the amplitude of the state is allowed to be a function.
It is also true for the case where the solutions to the Hamiltonian and ${\cal V}$
constraint are inconsistent, i.e. at finite $\gamma$. It would be interesting to identify the full non-perturbative state corresponding
to the solution (\ref{kodbcg1}) for the Euler-Pontryagin theory.

We close by noting that $Y_{CS}$ is only imaginary in MSS if $c=0$. Hence, even in MSS we can test the non-triviality of 
(\ref{realKodb}).

\end{document}